\documentclass[twocolumn]{aastex61}

\shorttitle{Limiting brightness temperature and Doppler factors for blazars}
\shortauthors{I. Liodakis}
\begin{document}

\title{Constraining the limiting brightness temperature and Doppler factors for the largest sample of radio bright blazars}

\correspondingauthor{I. Liodakis}
\email{ilioda@stanford.edu}

\author{Ioannis Liodakis}
\affil{KIPAC, Stanford University, 452 Lomita Mall, Stanford, CA 94305, USA}

\author{Talvikki Hovatta}
\affil{Tuorla Observatory, Department of Physics and Astronomy, University of Turku, V\"ais\"al\"antie 20, 21500 Kaarina, Finland}

\author{Daniela Huppenkothen}
\affil{Dirac Institute, University of Washington, Physics and Astrophysics Bldg, 3910 15th ave. NE Seattle, WA 98195-0002 }

\author{Sebastian Kiehlmann}
\affil{Owens Valley Radio Observatory, California Institute of Technology, Pasadena, CA 91125, USA}

\author{Walter Max-Moerbeck}
\affil{Universidad de Chile, Departamento de Astronom\' ia, Camino El Observatorio 1515, Las Condes, Santiago, Chile}

\author{Anthony C. S. Readhead}
\affil{Owens Valley Radio Observatory, California Institute of Technology, Pasadena, CA 91125, USA}

\begin{abstract}
Relativistic effects dominate the emission of blazar jets complicating our understanding of their intrinsic properties. Although many methods have been proposed to account for them, the variability Doppler factor method has been shown to describe the blazar populations best. We use a Bayesian hierarchical code called {\it Magnetron} to model the light curves of 1029 sources observed by the Owens Valley Radio Observatory's 40-m telescope as a series of flares with an exponential rise and decay, and estimate their variability brightness temperature. Our analysis allows us to place the most stringent constraints on the equipartition brightness temperature i.e., the maximum achieved intrinsic brightness temperature in beamed sources which we found to be $\rm \langle T_{eq}\rangle=2.78\times10^{11}K\pm26\%$. Using our findings we estimated the variability Doppler factor for the largest sample of blazars increasing the number of available estimates in the literature by almost an order of magnitude. Our results clearly show that $\gamma$-ray loud sources have faster and higher amplitude flares than $\gamma$-ray quiet sources. As a consequence they show higher variability brightness temperatures and thus are more relativistically beamed, with all of the above suggesting a strong connection between the radio flaring properties of the jet and $\gamma$-ray emission. 
\end{abstract}


\section{Introduction}\label{intro}

Blazar jets are known to show extremely fast variability, boosted emission, and apparent superluminal motion of jet components. These, as well as other unique features seen in blazars, are due to the relativistic effects dominating the emission from the jet. The relativistic effects arise from the preferential orientation of the jet typically within $<$20 degrees from our line of sight \citep{Readhead1978,Blandford1979,Scheuer1979,Readhead1980}. Radio emission in blazars is produced by relativistic electrons accelerated in the magnetic field of the jet emitting synchrotron radiation. It is characterized by a flat spectrum from the centimeter up to in some cases millimeter wavelengths thought to be the result of the superposition of multiple synchrotron self-absorbed spectra. Although the radio emission is considered to be less variable than e.g., optical or $\gamma$-rays, it can still be exceptionally variable and  bright with some sources reaching radio core brightness temperatures of $>10^{13}$K (e.g., \citealp{Kovalev2016}). Since the intrinsic brightness temperature of a jet is expected to be of the order of $\sim5\times 10^{10}$K \citep{Readhead1994}, this would suggest that the jets continue to be highly relativistic on very large scales far from the supermassive black hole. Quantifying the beaming properties of the jets is then necessary in order to understand their energetics at large scales. These relativistic effects are quantified by the Doppler factor ($\delta$) which is a function of the velocity of the jet and the angle to the line of sight $\delta=[\Gamma(1-\beta\cos\theta)]^{-1}$, where $\Gamma$ is the Lorentz factor ($\Gamma=1/\sqrt{1-\beta^2}$), $\beta$ is the velocity of the jet in units of speed of light ($\rm \beta=u_j/c$) and $\theta$ is the viewing angle. The Doppler factor, although a crucial parameter in the blazar paradigm dictating all of the observed properties of blazars, is notoriously difficult to estimate since there is no direct method to measure either $\beta$ or $\theta$. For this reason, many indirect methods have been proposed in order to estimate $\delta$ which usually involve different energetic (e.g., \citealp{Ghisellini1993,Mattox1993,Fan2013,Fan2014}) and/or causality arguments (e.g., \citealp{Lahteenmaki1999-III,Hovatta2009,Jorstad2005,Jorstad2017}) or fitting the spectral energy distribution (SED, e.g., \citealp{Ghisellini2014,Chen2018}) of $\gamma$-ray emitting blazars. 

However, different methods often yield discrepant results due to either assumptions that do not hold or the wrongful application of the methods (see e.g., \citealp{Liodakis2017-II}). \cite{Liodakis2015-II} using blazar population models \citep{Liodakis2015,Liodakis2017-III} evaluated a number of these methods and found that the variability Doppler factor method \citep{Lahteenmaki1999-III,Hovatta2009} is the most accurate and can describe both flat spectrum radio quasar (FSRQ) and BL Lac object (BL Lacs) populations. The method is based on the assumption of equipartition between the energy density of the magnetic field and the energy density of the radiating particles, achieved at the peak of prominent flares, implying a characteristic intrinsic brightness temperature \citep{Kellermann1969,Singal1986,Readhead1994}. By comparing the intrinsic (equipartition, $\rm T_\mathrm{eq}$) to the highest observed brightness temperature one can estimate $\delta$. The drawback of the method is that it is limited by the cadence of the observations which sets a limit to the fastest observed flare and consequently a limit to the observed brightness temperature \citep{Liodakis2015-II}. 

In order to mitigate the effects of limited cadence, \cite{Liodakis2017} used multi-wavelength radio light curves in order to identify and track the evolution of flares throughout frequencies which allowed the authors to provide constrains on the variability brightness temperature and hence the Doppler factor of 58 sources. However, the number of blazars with simultaneous multi-wavelength radio light curves is extremely limited compared to single-frequency observations. Then, the only way to overcome the effects of limited cadence is through monitoring programs with sufficiently high cadence to resolve even the fastest flares in radio.

In this work, we explore the radio beaming properties of jets by analyzing the light curves of 1029 blazars and blazar-like sources using data from the Owens Valley Radio Observatory's (OVRO) 40-m blazar monitoring program \citep{Richards2011}. We focus on constraining the equipartition brightness temperature and the variability Doppler factors for the sources in our sample. In section \ref{sampl} we present the sample and tools of the analysis, in sections \ref{TB_var} and \ref{equipartition} we estimate the highest brightness temperature for the sources in our sample and use blazar population models in order to constrain $\rm T_\mathrm{eq}$. In section \ref{D_var} we estimate the variability Doppler factors, Lorentz factors, and viewing angles based on our results on $\rm T_\mathrm{eq}$, and finally in section \ref{conc} we discuss the findings of this work. We have assumed the standard $\rm \Lambda{CDM}$ cosmology with $\rm \Omega_m=0.27$, $\rm \Omega_\Lambda=1-\Omega_m$ and $\rm H_0=71$ ${\rm km \, s^{-1} \, Mpc^{-1}}$ \citep{Komatsu2009}.

\section{Sample \& Analysis}\label{sampl}
\begin{figure*}
\resizebox{\hsize}{!}{\includegraphics[scale=1]{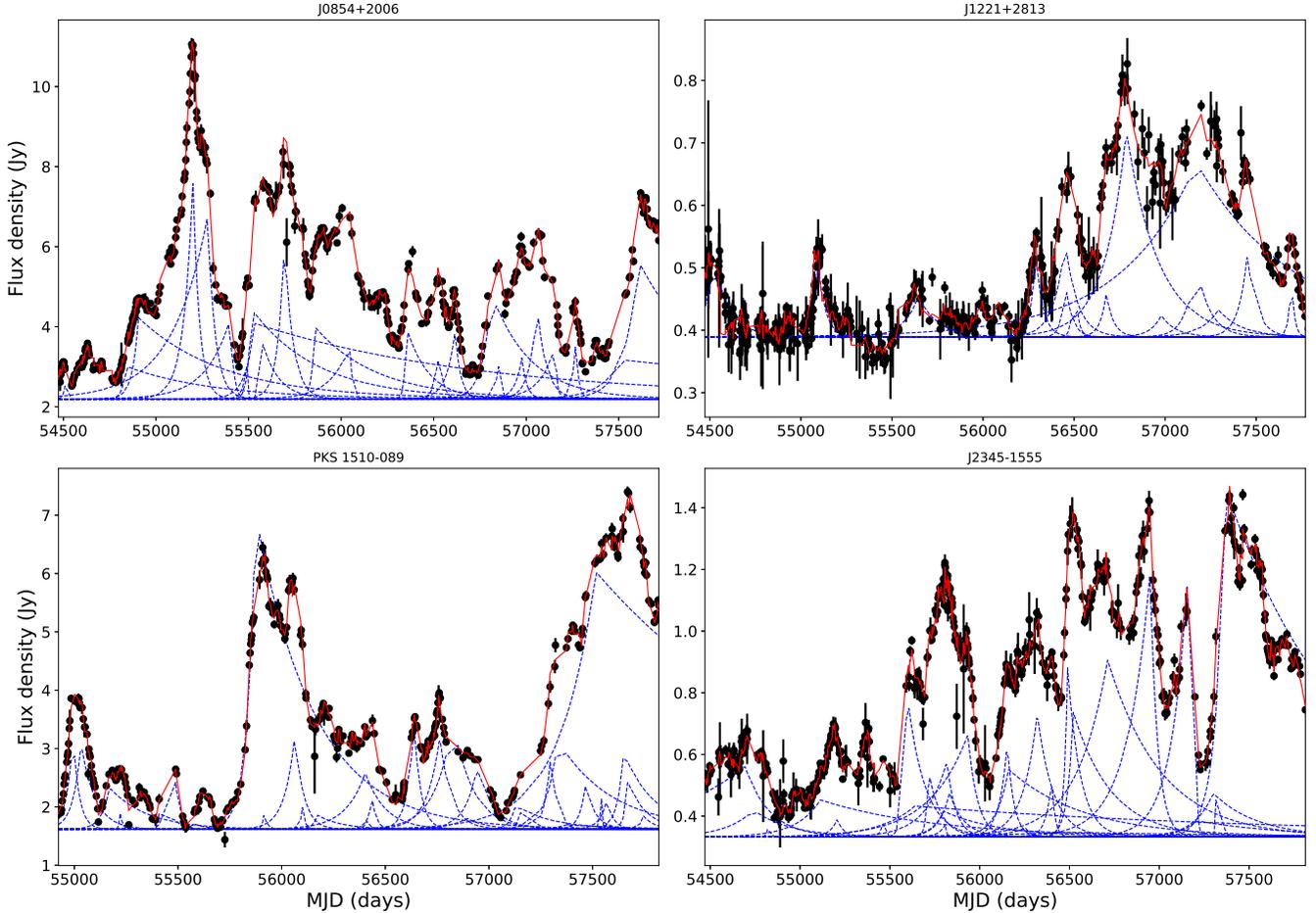} }
 \caption{Observed (black points) and posterior sampled (red lines) light curves for four sources in our sample namely J0854+2006 (upper left), J1221+2813 (upper right), PKS 1510-089 (lower left), and J2345-1555 (lower right). The blue dotted lines show the individual flares of one randomly selected realization of the light curve having added the background.}
\label{plt:example}
\end{figure*}
\begin{figure}
\resizebox{\hsize}{!}{\includegraphics[scale=1]{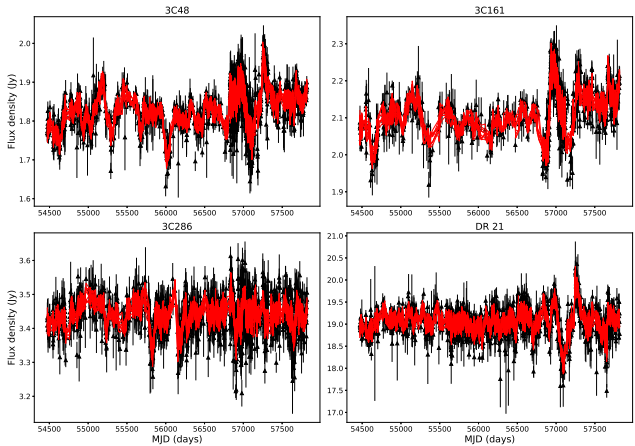} }
 \caption{Observed (black points) and posterior sampled (red lines) light curves for the four calibrator sources used by OVRO namely 3C46 (upper left), 3C161 (upper right), 3C286 (lower left), and DR 21 (lower right).}
\label{plt:example2}
\end{figure}

From the OVRO monitored sources ($\sim$1800), we selected those that showed prominent flares at 15~GHz via visual inspection of the light curves. Our final sample consists of 837 blazars (670 FSRQs, 167 BL Lacs) 58 radio galaxies and 134 yet unclassified sources, a total of 1029 sources. Since 2008, OVRO has been monitoring blazars in support of the {\it Fermi} $\gamma$-ray space telescope producing the most densely sampled radio light curves available to date with a cadence of about 3 days. The OVRO dataset provides the ideal opportunity to study the flaring and beaming properties of blazars since: (1) the unprecedented high cadence is able to resolve even the fastest flares in the most variable sources; (2) the light curves of most sources are sufficiently long (8-10 years) to include at least a few major events in each source.

For the analysis of the radio light curves we used {\it Magnetron} \citep{Huppenkothen2015}. {\it Magnetron} is a Bayesian hierarchical model implemented in {\it python} that models the light curves as a superposition of flares characterized by an exponential rise and exponential decay on top of a stochastic background. The shape of the flares is allowed to vary (the ratio of rise to decay time can be different in each flare) and the number of fitted flares in a light curve is a free parameter. Each flare is characterized by four parameters namely position, amplitude (in Jy), rise time (in days), and skewness (sk, decay/rise time ratio). The amplitude of a flare is defined as the difference between the peak flux density and the background level (see Figure 2 in \citealp{Huppenkothen2015}). The priors for the flare amplitudes and rise times are exponential while the priors for the skewness and the number of flares are uniform  distributions. The mean of the prior amplitude distribution takes values between [$10^{-10}$,150] Jy while the minimum and maximum of the uniform prior distribution for number of flares is [4,100]. All the prior distributions and their hyperparameters used by {\it Magnetron} are listed in Table 1 of \cite{Huppenkothen2015}.  In this work, contrary to \citealp{Huppenkothen2015}, we treat the background with a stochastic model (Ornstein--Uhlenbeck (OU) process) to account for intrinsic blazar variability not related to flaring events. The OU process is a stochastic, stationary Gauss--Markov process often used to treat AGN variability (e.g., \citealp{Kelly2009}). The version of {\it Magnetron} used in this work includes two new parameters to parametrize this stochastic process. The first quantity is the rate of mean reversion ($\alpha_{OU}$) which is included in the model through a parameter $\rm L$ as $\alpha_{\rm OU} = \exp{\rm (-1/L)}$. The prior for $\rm L$ is a log-uniform distribution such that $log(\rm L) \sim \rm Uniform(0.01*T, 0.01*T+1000)$, where $\rm T$ is the total length of the light curve. The second parameter is the volatility of the OU process, $\sigma_{\rm OU}$ i.e., the average magnitude per square root of time of random Brownian fluctuations. The prior for $\sigma_{\rm OU}$ is also log-uniform, such that $\log{\sigma_{\rm OU}}\sim\rm Uniform(10^{-3}, 10^{3})$. While previous attempts of fitting radio light curves used a constant value for the background level, using the OU process results in a varying background across the light curve. We have verified that using a different background model (such as a constant background or a simple random walk model) results in $\leq$10\% difference in the derived brightness temperatures, thus the choice of the background model does not affect our results in any significant way. The joint posterior probability distribution for the number of flares, and the parameters of all flares as well as the hyperparameters describing the distributions of flares are sampled using Diffusive Nested Sampling \citep{Brewer2009,Brewer2016}\footnote{https://github.com/eggplantbren/DNest4} allowing for a better exploration of the parameter space. Once the code has converged to the ``true'' posterior distribution, it samples $\sim10^2$ sets of flare parameters. These sets are different realizations of the flares in the observed light curve taking into account the inherent uncertainty in the parameters of the flares as well as the uncertainty in the number of flares of each light curve.  A more detailed description of {\it Magnetron} can be found in \cite{Huppenkothen2015} while the code is publicly available online on GitHub\footnote{https://github.com/dhuppenkothen/magnetron2/tree/blazars}.

Figure \ref{plt:example} shows the results of the light curve modeling for four sources in our sample as well as individual flares for one posterior sample in each source. All the light curves were visually inspected to ensure the simulated light curves are not affected by either spurious events or observational artifacts. In addition, we compared the rise times and amplitudes of the identified flares to test whether we were able to resolve, to OVRO's sensitivity, all the flaring events. For all classes of sources there is a lack of flares with high amplitudes and rise times close to the cadence of OVRO ($\sim 3$ days). In the case of FSRQs and unclassified sources we detect a mild positive correlation between the rise times and amplitudes according to the Spearman correlation test (Spearmann $\rho\approx0.3$, p-value $<10^{-5}$ for both classes). For all sources, we find that for rise times $<14$ days the majority of flares (60\%) have amplitudes lower than the median amplitude of the flares in the light curve. Out of the flares that have higher amplitudes than the median, less than 20\% have amplitudes higher than half of the maximum amplitude in the light curve. When we consider individual populations we find similar percentages ($\pm5\%$-$10\%$). All the above show that OVRO's cadence allowed us to resolve all the most significant events within the time span of the observations. A third quality test was to assess whether {\it Magnetron} is overfitting the data i.e., needlessly increasing the number of flares in a model to account for even the lowest flux-density variations, a common problem in usually employed $\chi^2$ fitting routines. We attempt to fit four sources used in the calibration of the OVRO observations were any flux-density fluctuations in the light curve are expected to be dominated by noise rather than any flaring activity.  Figure \ref{plt:example2} shows the observed and posterior sampled light curves for those calibrator sources. Although one could visually ``detect'' a number of apparent flares in each source (Fig.  \ref{plt:example2}), no more than two flares were detected by {\it Magnetron} in any given source for any given posterior sample of the light curves. This would suggest that the stochastic model for the background is able to adequately take into account the intrinsic low amplitude variability.

\section{Variability Brightness temperature}\label{TB_var}
\begin{figure}
\resizebox{\hsize}{!}{\includegraphics[scale=1]{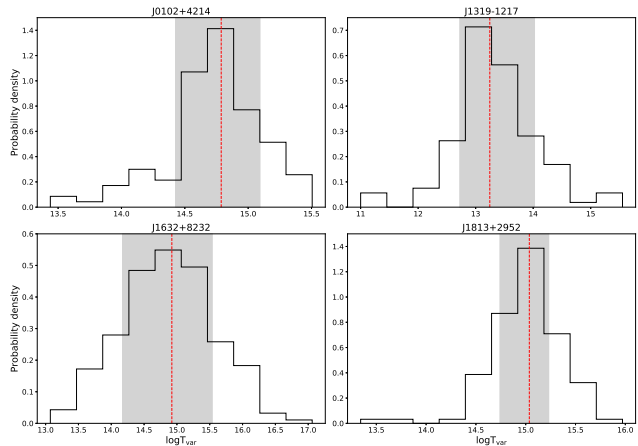} }
 \caption{Distribution of the logarithm of the maximum $\rm T_\mathrm{var}$ for four sources in our sample namely J0102+4214 (upper left), J1319-1217 (upper right), J1632+8232 (lower left), and J1813+2952 (lower right). The red dashed line shows the median and the grey shaded areas the 1$\sigma$ confidence internals in each source. }
\label{plt:tvar_unc}
\end{figure}
For every source in our sample with an available redshift estimate (all FSRQs, $\sim$71\% BL Lacs, $\sim$56\% radio galaxies and $\sim$42\% of unclassified sources), we estimate the variability brightness temperature ($\rm T_\mathrm{var}$) using,
\begin{equation}
\rm T_\mathrm{var}=1.47 \cdot 10^{13}\frac{d^2_L \Delta S_\mathrm{ob}(\nu)}{\nu^2t^2_\mathrm{var}(1+z)^4}~K,
\label{eq:tvar_num}
\end{equation}
where $\rm z$ is the redshift, $\rm \Delta S_\mathrm{ob}(\nu)$ the amplitude of the flare in Jy, $\rm d_L$ is the luminosity distance in Mpc, $\nu$ the observing frequency in GHz, and $\rm t_\mathrm{var}$ the rise time of a flare in days \citep{Liodakis2017}.  We calculate $\rm T_\mathrm{var}$ for every flare in a given posterior sample and find the maximum $\rm T_\mathrm{var}$ since that provides the strongest constrain on $\rm T_\mathrm{eq}$. We repeat the above process for all available samples (157 models on average) and create a distribution for $\rm T_\mathrm{var,max}$ for a given source. From that distribution we calculate the median and 1$\sigma$ confidence intervals which we quote as the uncertainty on $\rm T_\mathrm{var,max}$. Figure \ref{plt:tvar_unc} shows four examples of the maximum $\rm T_\mathrm{var}$ distributions. It is possible for the distributions to be narrower or wider than the ones shown in Fig. \ref{plt:tvar_unc}. The width of the distribution reflects on the ability of the modeling procedure to constrain the flare parameters' posterior distributions given the dataset. Thus the size of the confidence intervals of the $\rm T_\mathrm{var,max}$ distribution give a sense of how well we can constrain $\rm T_\mathrm{var,max}$ in that source. For simplicity we refer to  $\rm \langle T_\mathrm{var,max} \rangle$ as $\rm T_\mathrm{var}$ hereafter.

\begin{figure}
\resizebox{\hsize}{!}{\includegraphics[scale=1]{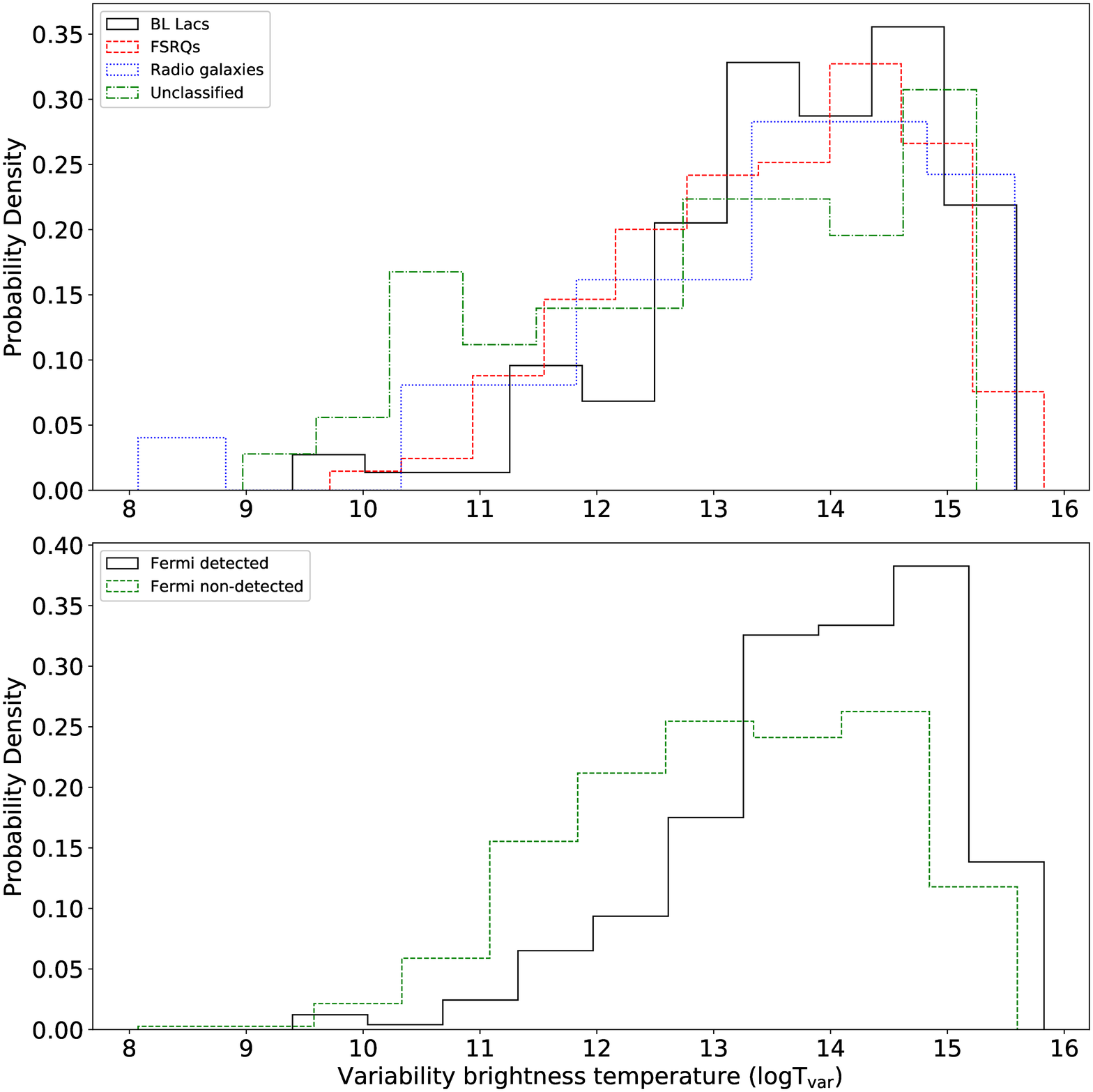} }
 \caption{{\bf Upper panel}: Logarithm of the maximum $\rm T_\mathrm{var}$ for BL Lacs (black solid), FSRQs (red dashed), radio galaxies (blue dotted), unclassified sources (green dashed-dotted). {\bf Lower panel}: Logarithm of the maximum $\rm T_\mathrm{var}$ for {\it Fermi} detected (black solid) and {\it Fermi} non-detected (green dashed) sources.}
\label{plt:tvar}
\end{figure}

Figure \ref{plt:tvar} shows the distribution of $\rm T_{var}$ for the different populations in our sample. The lowest brightness temperature ($\sim 10^8$K) is detected in a radio galaxy (M 81) while the highest ($6.7\times10^{15}$) is in a FSRQ (J0449+1121). There is only a marginal difference between the $\rm T_\mathrm{var}$ distributions of the BL Lacs and FSRQs according to the Wilcoxon rank-sum test (WRS test\footnote{The Wilcoxon rank-sum or Mann-Whitney U test operates under the null hypothesis that the two distributions are drawn from the same sample while the alternate hypothesis is that one sub-sample has systematically higher values than the other. The p-value threshold we are using is 0.05.}, p-value 0.0497) with BL Lacs having on average higher values. BL Lacs and FSRQs also show on average higher values than the unclassified sources (WRS p-value 0.005 and 0.028 respectively). No other significant difference between the distributions of the different populations was detected. It would be interesting to also separately compare the flaring properties of the sources (i.e., individually comparing flare amplitude (maximum) and flare rise time (shortest) distributions). The comparison between the different populations in our sample showed that although there is no significant difference in the flare amplitudes between BL Lacs and FSRQs (WRS p-value 0.76, median $\approx0.23$~Jy for both populations), BL Lac flares evolve on faster timescales (WRS p-value 0.00003, median $\approx9$ days compared to median $\approx17$ days for FSRQs). Radio galaxies and unclassified sources have on average lower flare amplitudes (WRS p-value $<0.0002$, median of $\approx0.18$ for both classes) than blazars, but their flare rise times are comparable to BL Lacs.

Another interesting comparison would be between $\gamma$-ray loud and $\gamma$-ray quiet sources. We separate our sample using the ROMA-BZCAT catalogue of known blazars (which is based on the 1FGL and 2FGL catalogues, \citealp{Massaro2009,Massaro2015}) according to whether a source has been detected by  {\it Fermi} i.e., a source showing $\gamma$-ray emission. We find that {\it Fermi} detected (382) sources have systematically higher values that the {\it Fermi} non-detected (496) sources (WRS p-value $\sim10^{-18}$,  median $1.27\times10^{14}$K for detected and $1.56\times10^{13}$K for non-detected sources, Fig. \ref{plt:tvar} bottom panel). Additionally, we compare their flaring properties as above. The comparison showed that {\it Fermi} detected sources flare on faster timescales, and have higher amplitude flares than non-detected sources (WRS p-value $<0.0001$ in both cases).

\section{Equipartition brightness temperature}\label{equipartition}
\begin{deluxetable}{cccc}
\tablenum{1}\label{tab:chi2}
\tablecaption{Parameter values for the best-fit normal $\rm T_\mathrm{eq}$ distribution for different flux-limits for the FSRQ population.}
\tablehead{\colhead{Flux-limit} & \colhead{Mean}  & \colhead{Standard deviation} & \colhead{ reduced-$\chi^2$}    }
\startdata
0.5~Jy & $4.72\times10^{11}$&  $8.7\times10^{10}$ & 0.08 \\
1.0~Jy &$3.65\times10^{11}$ & $4.0\times10^{10}$ & 0.07 \\
1.5~Jy &$2.78\times10^{11}$ & $7.2\times10^{10}$ & 0.04   \\
\enddata
\end{deluxetable}

In order to constrain $\rm T_\mathrm{eq}$ we use blazar population models \citep{Liodakis2017-III}. The population models are optimized using only the apparent velocity and redshift distributions from the MOJAVE survey \citep{Lister2005}, and can yield Doppler factor distributions within flux-limited samples independent of the assumption of equipartition. We define three flux-limited samples (0.5~Jy, 1~Jy and 1.5~Jy) above the nominal flux limit of the OVRO monitoring program (0.354~Jy) using the overall mean flux density of each source \citep{Liodakis2017-IV}. This allows us to assess how sensitive are our results to a given flux-limit. Using the population models we generate Doppler factor distributions for BL Lacs and FSRQs for every flux-limit. From Eq. \ref{eq:tvar_num} the variability Doppler factor ($\delta_\mathrm{var}$) is defined as, 
\begin{equation}
\rm \delta_\mathrm{var}=(1+z)\sqrt[3]{\frac{T_\mathrm{var}}{T_\mathrm{eq}}}.
\label{eq:varia-Doppler}
\end{equation}
We assume that $\rm T_\mathrm{eq}$ has a known distribution. We construct the $\rm T_\mathrm{var}$ distribution of the sample under consideration using the estimated $\rm T_\mathrm{var}$ for each source in that sample. We then use {Eq. \ref{eq:varia-Doppler} to derive an observed Doppler factor distribution. Then, we constrain the parameters of the assumed  $\rm T_\mathrm{eq}$ distribution by minimizing the reduced $\chi^2$ between the expected (population model) and observed Doppler factor distributions. For the distribution of $\rm T_\mathrm{eq}$ we tested a delta function, and normal, log-normal and uniform distributions with a parameter space [$\rm 10^{10}K-10^{13}K$]. Once the best-fit parameters of each distribution were determined, we used the Bayesian Information Criterion (BIC) to select the most suitable model for $\rm T_\mathrm{eq}$.

For FSRQs we find that the best model for $\rm T_\mathrm{eq}$ is a normal distribution for all three flux-limited samples we considered with very similar mean ($\mu$) and standard deviation ($\sigma$, Table \ref{tab:chi2}). All the other distributions that were tested (although yielded worse models according to BIC) converged to the same range of $T_\mathrm{eq}$ values. Although the results of the minimization for the different flux-limits are consistent, the 1.5~Jy sample is the flux-limit to which the population models have been optimized, and thus where their strength lies (see discussion in \citealp{Liodakis2015}). For this reason we adopt the results from the 1.5~Jy sample for the FSRQs.

For BL Lacs, we also find that the best-fit distribution is normal for all flux-limits, however, the parameters of the inferred distributions all significantly exceed the inverse-Compton catastrophe limit ($10^{12}$K, \citealp{Kellermann1969}). Since we have yet to observe the extreme behavior predicted by the inverse-Compton catastrophe such a scenario is unlikely. A possible explanation is that the maximum Doppler factor inferred for BL Lacs by the population models is $\delta\approx30$ (the maximum $\delta$ in FSRQs is $\delta\approx60$) which given the high variability brightness temperatures seen in BL Lacs forces the very high $\rm T_\mathrm{eq}$. It is discussed in  \cite{Liodakis2017-III} that the BL Lac population ($\sim$16 sources) in the MOJAVE 1.5 Jy flux-limited sample, to which the population models are optimized, might not be a representative sample of BL Lacs, but rather a biased subsample of the brightest BL Lacs at 15~GHz. Hence, the population models cannot adequately describe the entirety of the BL Lac population present in our sample. Given that equipartition is determined by the jet processes and synchrotron physics, we expect the value of $\rm  T_\mathrm{eq}$ to be fairly similar for the different supermassive black hole powered jets. Thus we adopt the results from the FSRQs for all the populations in our sample ($\rm \langle T_{eq}\rangle=2.78\times10^{11}K\pm26\%$).

\section{Variability Doppler factors}\label{D_var}

\begin{figure}
\resizebox{\hsize}{!}{\includegraphics[scale=1]{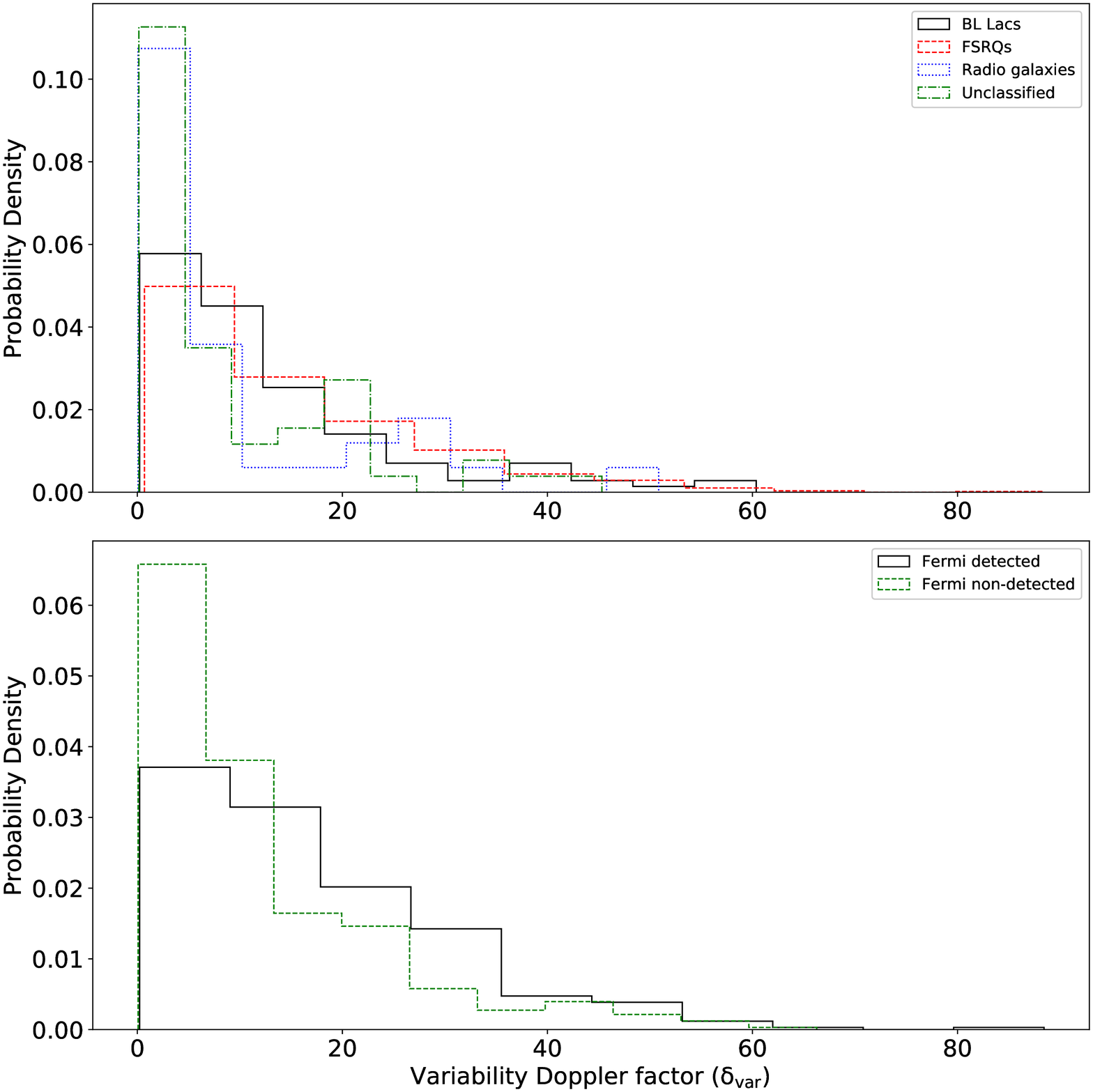} }
 \caption{{\bf Upper panel}: Variability Doppler factor distribution for BL Lacs (black solid), FSRQs (red dashed), radio galaxies (blue dotted), unclassified sources (green dashed-dotted). {\bf Lower panel}: Variability Doppler factor distribution for {\it Fermi} detected (black solid) and {\it Fermi} non-detected (green dashed) sources.}
\label{plt:dvar}
\end{figure}

In order to calculate $\delta_\mathrm{var}$,  we draw a random value from the $\rm T_\mathrm{var,max}$ distribution of each source and a random value for $\rm T_\mathrm{eq}$ from a Gaussian distribution with mean $\rm \langle T_{eq}\rangle=2.78\times10^{11}K$ and standard deviation  $\rm \sigma_{T_{eq}}=7.2\times10^{10}$.  Using Eq. \ref{eq:varia-Doppler} we calculate a $\delta_\mathrm{var}$. By repeating this process $10^3$ times   we create a distribution of $\delta_\mathrm{var}$ for every source. From the resulting $\delta_\mathrm{var}$ distribution of each source we estimate the median and 1$\sigma$ confidence intervals. Figure \ref{plt:dvar} shows the distribution of $\delta_\mathrm{var}$ for the different populations (top panel) and {\it Fermi} detected and non-detected sources (bottom panel). BL Lacs and FSRQs have median values of $\delta_{\rm var}\approx10$ and $\delta_{\rm var}\approx11$ respectively while radio galaxies and unclassified sources have median $\rm \delta_{var}\approx5$. As expected, blazars have systematically higher Doppler factors than radio galaxies (WRS p-value $\sim$0.03) and unclassified sources (p-value $<$0.0002). We find no significant difference between the $\rm \delta_{var}$ distributions of the blazar classes (BL Lacs and FSRQs, WRS p-value 0.08). Comparing {\it Fermi} detected and non-detected sources we found that {\it Fermi} detected sources have systematically higher values than non-detected sources (WRS p-value $\sim10^{-13}$, median $\rm \delta_{var}\approx14$ for detected and $\rm \delta_{var}\approx8$ for non-detected sources). This would suggest that the $\gamma$-ray emission could be, in part, resulting from the enhanced relativistic effects in these sources. 

There are 31 sources (10 FSRQs, 5 BL Lacs, 5 radio galaxies, 11 unclassified sources) that show $\rm \delta_{var}<1$, 7 of which have been detected by {\it Fermi} (4 BL Lacs, 2 radio galaxy, 1 unclassified source). These sources are either misaligned AGN with jets pointing away from our line of sight and thus their emission is de-boosted, or are mildly beamed and have not shown any radio bright flaring events with $\rm T_{\rm var}>10^{11}$K during the OVRO monitoring period (2008-2017).

\subsection{Lorentz factors \& Viewing angles}
\begin{figure}
\resizebox{\hsize}{!}{\includegraphics[scale=1]{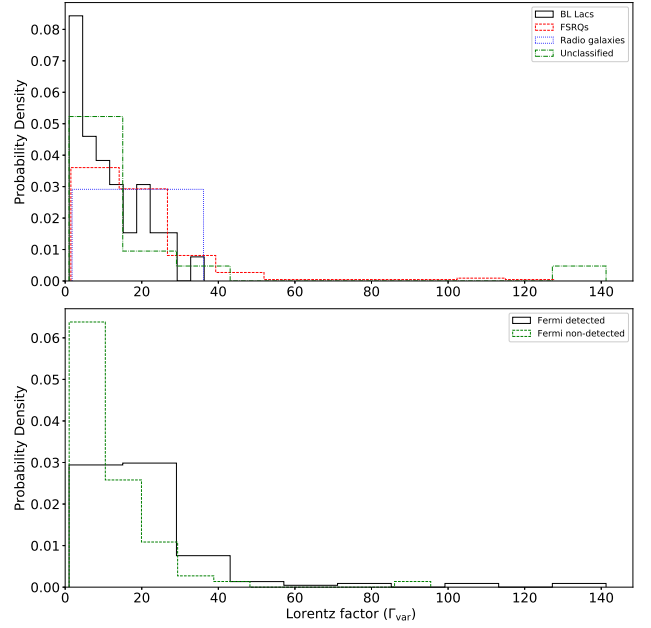} }
 \caption{{\bf Upper panel}: Lorentz factor distribution for BL Lacs (black solid), FSRQs (red dashed), radio galaxies (blue dotted), unclassified sources (green dashed-dotted). {\bf Lower panel}: Lorentz factor distribution for {\it Fermi} detected (black solid) and {\it Fermi} non-detected (green dashed) sources.}
\label{plt:gvar}
\end{figure}
\begin{figure}
\resizebox{\hsize}{!}{\includegraphics[scale=1]{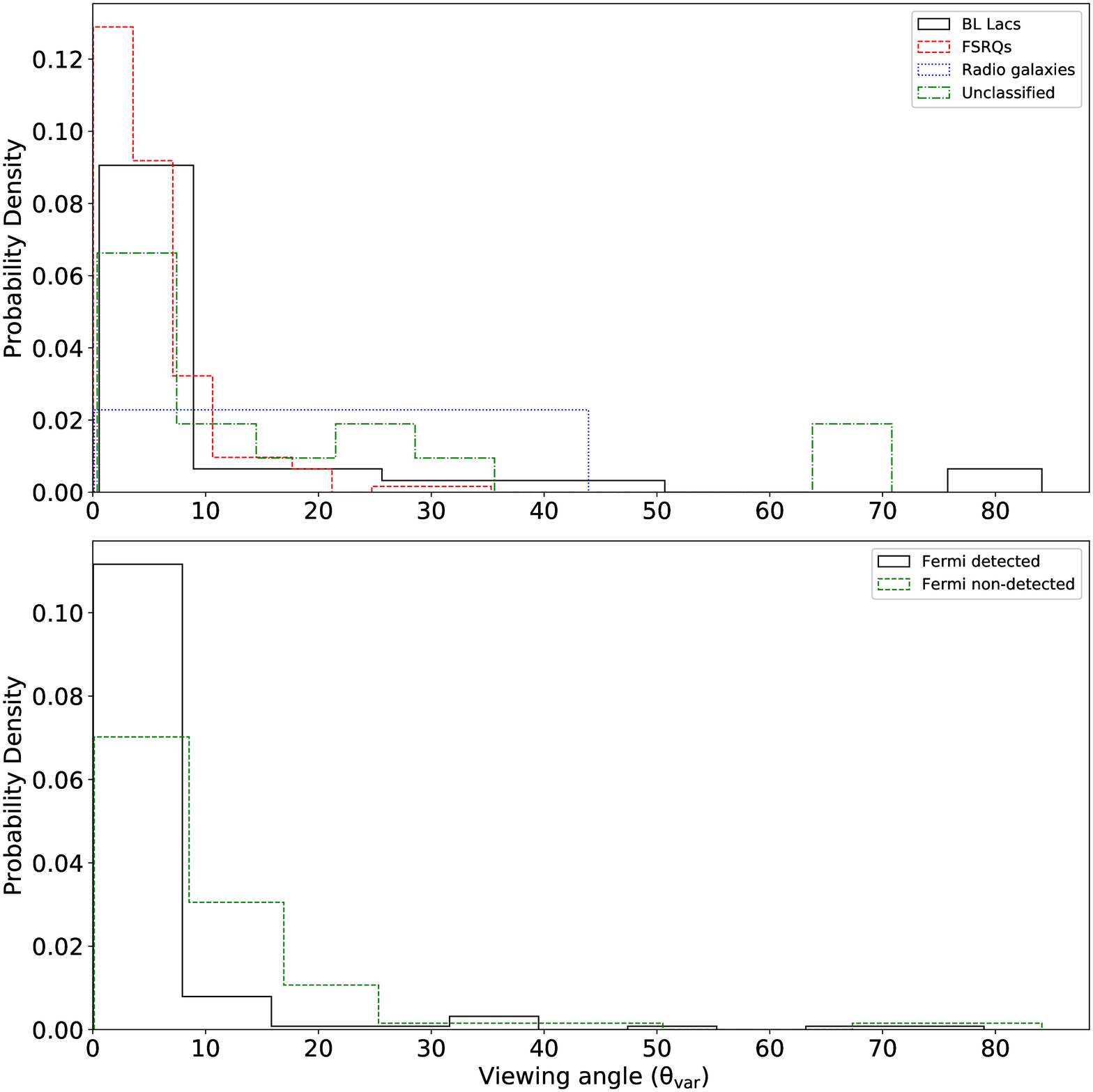} }
 \caption{{\bf Upper panel}: Viewing angle distribution for BL Lacs (black solid), FSRQs (red dashed), radio galaxies (blue dotted), unclassified sources (green dashed-dotted). {\bf Lower panel}: Viewing angle distribution for {\it Fermi} detected (black solid) and {\it Fermi} non-detected (green dashed) sources.}
\label{plt:thvar}
\end{figure}
\begin{figure}
\resizebox{\hsize}{!}{\includegraphics[scale=1]{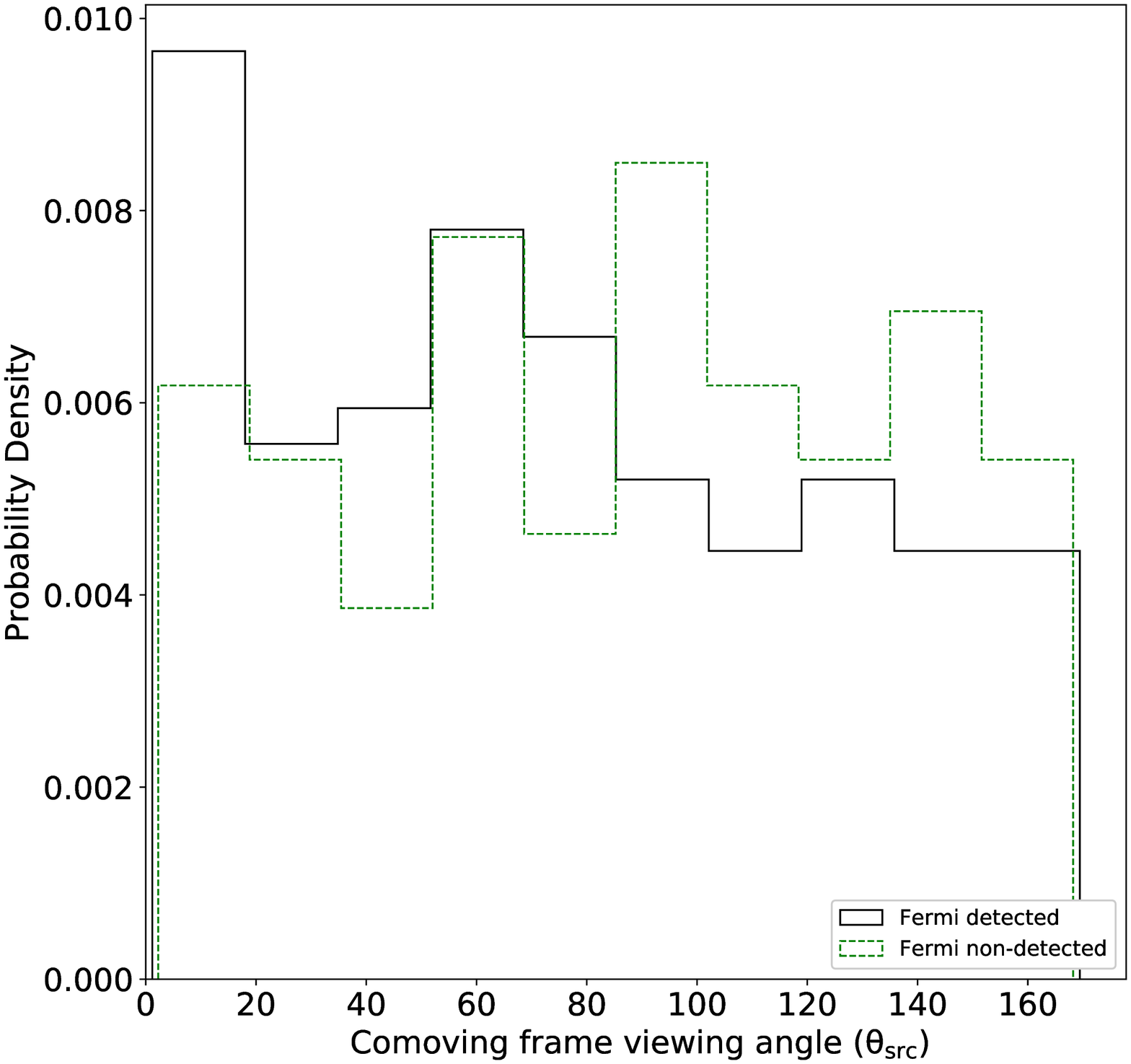} }
 \caption{Comoving frame viewing angle distribution for {\it Fermi} detected (black solid) and {\it Fermi} non-detected (green dashed) sources.}
\label{plt:thvar_src}
\end{figure}

Using the apparent velocity of the resolved jet components we can estimate both $\Gamma$ and $\theta$ as,
\begin{equation}
\rm  \Gamma_\mathrm{var} = \frac{\beta_\mathrm{app}^2 + \delta_\mathrm{var}^2 + 1}{2\delta_\mathrm{var}},
\label{eq:lorentz}
\end{equation}
\begin{equation}
\rm \theta_\mathrm{var} = \arctan \left(\frac{2\beta_\mathrm{app}}{\beta_\mathrm{app}^2 + \delta_\mathrm{var}^2-1}\right),
\label{eq:theta}
\end{equation}
where $\beta_\mathrm{app}$ is the apparent velocity. For $\beta_\mathrm{app}$ we use data from the MOJAVE survey  \citep{Lister2005,Lister2016}. For our calculations we use the maximum observed apparent velocity in each jet. There are 238 sources with an available estimate 160 of which have been detected by {\it Fermi}. Figures \ref{plt:gvar} and \ref{plt:thvar} show the Lorentz factor and viewing  angle distributions for the different classes (top panels) and {\it Fermi} detected and non-detected sources (bottom panels). From the sources with an apparent velocity measurement, FSRQs have  on average faster jets than any other source class (WRS p-value $<$0.006, median $\Gamma_\mathrm{var}\approx15.6$ compared to about 10 for BL Lacs, 6.3 for radio galaxies, and 7.2 for unclassified sources). Similarly {\it Fermi} detected have on average faster jets ($\sim10^{-6}$, median $\Gamma_\mathrm{var}\approx17$) than {\it Fermi} non-detected sources (median $\Gamma_\mathrm{var}\approx9$). There are four sources (namely J0108+0135, J0948+4039, J1613+3412, and C1724+4004) with a high Lorentz factor ($\Gamma_\mathrm{var}>100$). The derived Doppler factors for these sources are $<5$ yet the measured $\beta_\mathrm{app,max}$ are $>20$. Using the median $\beta_\mathrm{app}$ instead of $\beta_\mathrm{app,max}$ brings the $\Gamma_\mathrm{var}$ estimates to much lower values ($<50$). However, in all cases the components that yielded $\beta_\mathrm{app,max}$ in each source were ejected prior to the beginning of the observations considered here. It is then possible that a major flaring event not considered in this work is associated with these components, and hence for these sources we are underestimating their Doppler factors. BL Lacs and FSRQs have similar viewing angles (WRS p-value 0.13, median $\theta_\mathrm{var}\approx4$ for BL Lacs and median $\theta_\mathrm{var}\approx5$ for FSRQs) while radio galaxies and unclassified sources have on average larger (p-value $<$0.04, median of 12 degrees, and 9 degrees respectively). {\it Fermi} non-detected sources also have on average larger viewing angles (p-value $\sim10^{-9}$, median of 7.2 degrees) compared to {\it Fermi} detected sources (median of 2.7 degrees).

A similar comparison of the beaming properties for a smaller sample (62 sources in total) in \cite{Savolainen2010} found that there is a lack of small viewing angles in the comoving frame of the jet for {\it Fermi} detected sources. The comoving-frame viewing angle ($\theta_{src}$) is defined as,
\begin{equation}
\rm \theta_\mathrm{src} = \arccos \left( \frac{\cos\theta_{\rm var}-\beta}{1-\beta\cos\theta_{\rm var}}\right).
\label{eq:theta_src}
\end{equation}
Based on the fact that {\it Fermi} detected sources have on average higher apparent velocities \citep{Lister2009-III} and VLBI core brightness temperatures \citep{Kovalev2009}, the authors concluded that if the lack of small $\rm \theta_\mathrm{src}$ persists in a larger sample would suggest either anisotropy of the rest-frame $\gamma$-ray emission or a dependence of that emission on the Lorentz factor. Our sample allows us to test whether such lack of small $\rm \theta_\mathrm{src}$ exists. Figure \ref{plt:thvar_src} shows the $\rm \theta_\mathrm{src}$ distributions for {\it Fermi} detected and non-detected sources. With the significantly larger sample considered in this work we do not find any lack of small $\rm \theta_\mathrm{src}$ for {\it Fermi} detected sources. Thus our results disfavor scenarios involving any dependence on the Lorentz factor or anisotropic rest-frame $\gamma$-ray emission.

Table \ref{tab:var_prop} lists the values for $\rm  T_\mathrm{var}$, $\delta_{\rm var}$, $\Gamma_{\rm var}$, and $\theta_{\rm var}$ and their uncertainties. For all the sources (with or without a $\beta_{\rm app,max}$ estimate) we use the $\beta_{\rm app,max}$ distribution to bracket the possible range of $\Gamma_{\rm var}$, and $\theta_{\rm var}$ estimates for that source given the estimated $\delta_{\rm var}$.

%\onecolumn

\begin{deluxetable*}{ccccccccccccccccc}
\tablenum{2}\label{tab:var_prop}
\tablecaption{Variability brightness temperatures and beaming properties for the sources in our sample}
\tablehead{\colhead{Name} & \colhead{Class}  & \colhead{z}& \colhead{$\beta_{\rm app.max}$}& \colhead{$\sigma_{\beta_{\rm app,max}}$} & \colhead{$\rm T_{\rm var}$} & \colhead{-$\rm \sigma_{T_{\rm var}}$} & \colhead{$\rm \sigma_{T_{\rm var}}$} & \colhead{$\delta_{\rm var}$} & \colhead{-$\sigma_{\delta_{\rm var}}$} & \colhead{$\sigma_{\delta_{\rm var}}$}  & \colhead{$\Gamma_{\rm var}$}& \colhead{$\Gamma_{\rm min}$}& \colhead{$\Gamma_{\rm max}$} & \colhead{$\theta_{\rm var}$}& \colhead{$\theta_{\rm min}$}& \colhead{$\theta_{\rm max}$} }
\startdata
J0001-1551 & F & 2.044 & - & - & 11.15 & -1.12 & 0.64 & 2.51 & -1.44 & 1.69 & - & 1.45 & $>100$ & - & 0.91 & 23.47 \\ 
J0001+1914 & F & 3.100 & - & - & 10.29 & -0.25 & 3.64 & 1.82 & -0.5 & 30.76 & - & 1.19 & $>100$ & - & 2.07 & 33.24 \\ 
J0004+2019 & B & 0.677 & - & - & 11.14 & -0.52 & 3.83 & 1.37 & -0.48 & 24.59 & - & 1.05 & $>100$ & - & 2.88 & 47.06 \\ 
J0004+4615 & F & 1.810 & - & - & 12.78 & -0.63 & 0.2 & 7.75 & -2.75 & 1.7 & - & 3.94 & $>100$ & - & 0.08 & 7.42 \\ 
J0005+3820 & F & 0.229 & - & - & 13.26 & -0.31 & 0.63 & 5.23 & -1.22 & 2.72 & - & 2.71 & $>100$ & - & 0.18 & 11.02 \\ 
J0006-0623 & B & 0.347 & 7.31 & 0.33 & 13.55 & -0.5 & 0.77 & 6.96 & -2.27 & 5.63 & 7.39 & 3.55 & $>100$ & 8.25 & 0.1 & 8.26 \\ 
J0010+1058 & - & 0.089 & 1.58 & 0.29 & 12.48 & -0.08 & 1.38 & 2.51 & -0.3 & 4.24 & 1.95 & 1.45 & $>100$ & 22.1 & 0.91 & 23.51 \\ 
J0010+1724 & F & 1.601 & - & - & 14.38 & -0.48 & 0.45 & 24.81 & -7.46 & 12.21 & - & 12.43 & 44.18 & - & 0.01 & 2.31 \\ 
J0010+2047 & F & 0.600 & - & - & 13.18 & -1.46 & 0.45 & 6.02 & -4.05 & 3.12 & - & 3.09 & $>100$ & - & 0.14 & 9.56 \\ 
J0011+0057 & F & 1.492 & - & - & 13.34 & -1.37 & 1.92 & 11.03 & -7.63 & 33.64 & - & 5.56 & 76.98 & - & 0.04 & 5.2 \\  
\enddata
\tablecomments{Names are as listed in the OVRO website. The values of $\rm T_{\rm var}$ and its uncertainties are given in $\log_{10}$. -$\rm \sigma_{T_{\rm var}}$, $\rm \sigma_{T_{\rm var}}$ and -$\sigma_{\delta_{\rm var}}$, $\sigma_{\delta_{\rm var}}$ are the asymmetric uncertainties on $\rm T_{\rm var}$ and $\delta_{\rm var}$ respectively. [$\Gamma_{\rm min}$,$\Gamma_{\rm max}$] and [$\theta_{\rm min}$,$\theta_{\rm max}$] are the possible minimum and maximum values of each source for a given  $\delta_{\rm var}$ by marginalizing over the $\beta_{\rm app,max}$ distribution. The redshift estimates are taken from \cite{Richards2014}, SIMBAD \citep{Wenger2000}, NASA/IPAC Extragalactic Database (NED\footnote{NASA/IPAC Extragalactic Database (NED) is operated by the Jet Propulsion Laboratory, California Institute of Technology,\\ under contract with the National Aeronautics and Space Administration.}), and the MOJAVE database \citep{Lister2018}. The table lists only the first 10 sources. It is published in its entirety in the machine-readable format. A portion is shown here for guidance regarding its form and content.}
\end{deluxetable*}

\subsection{Sources without redshift}

There are 151 sources in our sample without an available redshift estimate. Out of these sources 49 are BL Lacs, 25 radio galaxies, and 77 are unclassified sources. We follow the same procedure for the sources with redshift and calculate $\rm T_\mathrm{var}$ using Eq. \ref{eq:tvar_num} without the cosmological correction. We use the minimum and maximum redshift estimates [0.00014,5.47] in our sample to calculate lower and upper limits for $\rm T_\mathrm{var}$, and $\delta_{\rm var}$ using the mean $\rm T_\mathrm{eq}$ derived in section \ref{equipartition}. Similarly, we use the $\beta_{\rm app,max}$ distribution to bound the possible $\Gamma_{\rm var}$, and $\theta_{\rm var}$ estimates for these sources. We list all those values in Table \ref{tab:var_prop_noz}.

\begin{deluxetable*}{ccccccccccc}
\tablenum{3}\label{tab:var_prop_noz}
\tablecaption{Variability brightness temperatures and beaming properties for the sources in our sample without a redshift estimate}
\tablehead{\colhead{Name} & \colhead{Class}  & \colhead{$\rm T_{\rm var,no-z}$} & \colhead{$\rm T_{\rm var,min}$} & \colhead{$\rm T_{\rm var,max}$} & \colhead{$\delta_{\rm var,min}$} & \colhead{$\delta_{\rm var,max}$} &\colhead{$\Gamma_{\rm var,min}$} & \colhead{$\Gamma_{\rm var,max}$}& \colhead{$\theta_{\rm var,min}$}& \colhead{$\theta_{\rm var,max}$}}
\startdata
J0004-1148 & B & 8.03 & 7.57 & 14.23 & 0.05 & 55.02 & 9.79 & $>100$ & 0.0 & 88.79 \\ 
J0009+0628 & B & 8.04 & 7.58 & 14.24 & 0.05 & 55.42 & 9.72 & $>100$ & 0.0 & 88.79 \\ 
J0019+2021 & B & 6.45 & 6.0 & 12.66 & 0.02 & 16.41 & 8.24 & $>100$ & 0.0 & 88.86 \\ 
J0022+0608 & B & 8.26 & 7.8 & 14.46 & 0.06 & 65.51 & 8.23 & $>100$ & 0.0 & 88.76 \\ 
J0035-1305 & - & 7.87 & 7.42 & 14.08 & 0.05 & 48.78 & 11.04 & $>100$ & 0.0 & 88.81 \\ 
J0105+4819 & - & 8.4 & 7.94 & 14.6 & 0.07 & 73.06 & 7.39 & $>100$ & 0.0 & 88.74 \\ 
J0106+1300 & - & 8.89 & 8.43 & 15.09 & 0.1 & 106.29 & 5.1 & $>100$ & 0.0 & 88.59 \\ 
J0112+2244 & B & 9.02 & 8.56 & 15.22 & 0.11 & 117.52 & 4.63 & $>100$ & 0.0 & 88.53 \\ 
J0132+4325 & - & 7.86 & 7.4 & 14.06 & 0.04 & 48.14 & 11.18 & $>100$ & 0.0 & 88.81 \\ 
J0202+4205 & B & 5.14 & 4.69 & 11.35 & 0.01 & 6.0 & 3.08 & $>100$ & 0.0 & 88.86 \\  
\enddata
\tablecomments{Names are as listed in the OVRO website. The values of $\rm T_{\rm var}$ are given in $\log_{10}$. Column (3) lists the $\rm T_{\rm var}$ estimate for each source without the cosmological correction ($d^2_L/(1+z)^4$, Eq. \ref{eq:tvar_num}). [$\Gamma_{\rm min}$,$\Gamma_{\rm max}$] and [$\theta_{\rm min}$,$\theta_{\rm max}$] are the possible minimum and maximum values of each source for the min and max $\delta_{\rm var}$ by marginalizing over the $\beta_{\rm app,max}$ distribution. The table lists only the first 10 sources. It is published in its entirety in the machine-readable format. A portion is shown here for guidance regarding its form and content.}
\end{deluxetable*}

\subsection{Comparison with other Doppler factor estimation methods}

\begin{deluxetable}{ccccc}
\tablenum{4}\label{tab:dvar_true}
\tablecaption{List of sources with Doppler factors consistent with J17 within 1$\sigma$.}
\tablehead{\colhead{OVRO name} & \colhead{J17 name} & \colhead{$\delta_{\rm var}$} & \colhead{$\delta_{\rm J17}$}}
\startdata
J0238+1636 & 0235+164 & 43.53$^{+19.79}_{-11.49}$ & 52.8 $\pm$ 8.4 \\  
J0339-0146 & 0336-019 & 23.09$^{+18.9}_{-6.08}$ & 15.7 $\pm$ 4.9 \\ 
0415+379 & 0415+379 & 1.99$^{+1.55}_{-0.43}$ & 2.0 $\pm$ 0.5 \\ 
J0433+0521 & 0430+052 & 4.16$^{+1.42}_{-1.09}$ & 4.5 $\pm$ 2.0 \\   
J0830+2410 & 0827+243 & 31.97$^{+5.83}_{-4.2}$ & 22.8 $\pm$ 8.5 \\ 
C1224+2122 & 1222+216 & 5.32$^{+7.84}_{-1.76}$ & 7.4 $\pm$ 2.1 \\ 
J1229+0203 & 1226+023 & 3.78$^{+1.1}_{-0.55}$ & 4.3 $\pm$ 1.3 \\ 
J1310+3220 & 1308+326 & 26.35$^{+13.39}_{-16.63}$ & 20.9 $\pm$ 1.2 \\ 
PKS1510-089 & 1510-089 & 32.14$^{+8.07}_{-7.97}$ & 35.3 $\pm$ 4.6 \\ 
J1751+0939 & 1749+096 & 17.62$^{+10.1}_{-3.16}$ & 17.7 $\pm$ 7.7 \\
J2253+1608 & 2251+158 & 26.61$^{+6.28}_{-2.97}$ & 24.4 $\pm$ 3.7 \\  
\enddata
\tablecomments{Names are as given in the OVRO website and J17.}
\end{deluxetable}

There are several methods in the literature for estimating the Doppler factor in blazar jets, some of which are mentioned in section \ref{intro}. Although a broader comparison study between the different methods similar to \cite{Liodakis2015-II,Liodakis2017-II} could be beneficial, we focus on recent results from the radio regime and SED modeling. The most recent attempts in estimating the variability Doppler factor for a large number of sources are \cite{Hovatta2009,Liodakis2017} (hereafter H09 and L17 respectively). In H09 the authors used data from the Mets\"ahovi monitoring program at 22 and 37~GHz \citep{Teraesranta1998} and estimated the variability brightness temperature for 87 sources by modeling the light curves using the same exponential rise and exponential decay model as this work. L17 used multi-wavelength radio data (2.64-142.33~GHz) from the F-GAMMA program \citep{Fuhrmann2016} to decompose the light curves using non-parametric models for the flare profiles tailored to the individual light curves of 58 sources. The cadence of H09 is weekly while the cadence for L17 is every two weeks to monthly. All of the sources in H09 and L17 are also in our sample. Our results appear to be consistent with both studies with roughly 50\% of the estimates consistent within 1$\sigma$. However, both H09 and L17 have assumed that $\rm T_\mathrm{eq}=5\times10^{10}$. Once we account for the different $\rm T_\mathrm{eq}$, the estimates derived in this work become larger by a factor of $\approx1.85$. The number of sources with consistent estimates drops to roughly 20\% and there is now a significant difference in the Doppler factor distributions with estimates of this work being systematically larger (WRS p-value $<0.0002$ for both samples). The higher Doppler factors from this work are most likely due to OVRO's faster cadence. However, cadence may not be solely responsible for the differences between the estimates. The dataset used in H09 includes observations up to roughly 2006. While there is overlap between the observing periods of L17 and this work, the estimates in both H09 and L17 originate from a variety of frequencies which may probe regions not co-spatial with the one probed at 15~GHz due to synchrotron self-absorption. It is then possible for the differences in the estimates to also be attributed to either significant flaring events have occurred outside the periods considered in H09 and L17 or that their reported estimates simply correspond to different regions of the jet. Additionally, results from the F-GAMMA survey would suggest a decreasing trend of the brightness temperature with frequency $\rm T_\mathrm{var}\propto\nu^{-1.2}$ \citep{Fuhrmann2016}. From Eq. \ref{eq:varia-Doppler} the Doppler factor should then decrease as $\rm \delta_\mathrm{var}\propto\nu^{-0.4}$ with increasing frequency. Such a trend could imply that the jets are accelerating from the higher to the lower radio frequencies which could explain some of the discrepancies. About 75\% of blazars in the MOJAVE survey have indeed shown at least one accelerating jet feature at 15~GHz \citep{Homan2015}. However, the fact that a significant fraction of the estimates in L17 are estimated at a lower frequency than 15~GHz would suggest that this scenario is unlikely to explain the discrepancies between the estimates from L17 and this work.

A more interesting comparison would be with the estimates in \cite{Jorstad2017} (hereafter J17). The method uses the variability timescales of individual jet components which are related to the Doppler factor through the observed angular size of the components derived from VLBI observations at 43~GHz \citep{Jorstad2005}. Although the method is also limited by the cadence of observations, it has the advantage of being independent of the assumption of equipartition. Thus agreement in the estimates of the two methods (J17 and this work) provides strong constrains for the Doppler factors of the jets. All of the sources (36) in J17 are included in the present sample. The estimates for 11/36 sources are consistent within 1$\sigma$. Differences in the estimates between the two samples are most likely attributed to the different assumptions used in each method or to reasons described above, however, no systematic difference is detected between the Doppler factor distributions according to the WRS test (p-value 0.56). The names and $\delta$ estimates of the sources in agreement between the two samples are given in Table \ref{tab:dvar_true}. 

SED modeling has also been used to constrain the $\rm \Gamma$ and $\rm \theta$ in blazar jets in part due to the fact that different $\gamma$-ray emission mechanisms are affected differently by the relativistic effects (e.g., \citealp{Dermer1995}). Recent SED modeling of a large number of sources found that the distribution of the derived Lorentz factors (a frequent assumption in SED modeling is that $\rm \delta=\Gamma$) is narrow, peaking at $\rm \delta=\Gamma\sim13\pm1.4$ \citep{Ghisellini2014}. Similar results were found in \cite{Chen2018} considering a larger sample ($\rm \delta=\Gamma\sim14$). It is usually assumed in SED modeling that the $\gamma$-ray emission is produced closer to the supermassive black hole than the radio core of the jet where most of the radio emission originates. It is then interesting that we find similar results for $\rm \delta_\mathrm{var}$ for the Fermi detected sources (median $\rm \delta_{var}\approx14$). The derived $\rm \Gamma_{var}$ in this work appears to be on average larger (median $\rm \Gamma_{var}\approx17$), however, the distribution is wide enough to prevent us from investigating any potential discrepancy. Agreement between the two methods could suggest that there is no significant change in the relativistic effects between the radio and $\gamma$-ray emission regions which has implications for the different jet acceleration scenarios as well as the possible location of the $\gamma$-ray production site. However, given the complexity of the SED models and the covariance between the different parameters involved in these models, any agreement could be artificial. A dedicated study of the sources studied in this work could allow us to probe possible differences in the relativistic effects between radio and $\gamma$-rays.

\section{Discussion \& Conclusions}\label{conc}

By modeling with a superposition of flares on top of a stochastic background the radio light curves from the OVRO 40-m telescope's blazar monitoring program we were able to  estimate the variability brightness temperatures and Doppler factors for 1029 sources, the largest set of estimates available to date. OVRO's high cadence allowed us to resolve even the fastest flares and set the strongest constrains on the highest $\rm T_{var}$ in each source. The present analysis is, however, limited by the time span of observations. It is possible for significant flaring events to have occurred outside the observing time-span considered here as the variability time scales in blazars are typically long \citep{Hovatta2007}. Thus, for all intents and purposes the results from this work should be treated as lower limits. The fact that roughly 12\% of our sources have Doppler factors as high as $\rm \delta_{var}>30$ would suggest that at least for a fraction of our sample we were able to estimate the ``true'' $\rm \delta_{var}$ of the jet. It would be productive to repeat such analysis with light curves observed during different time intervals (with similar cadence) than the one considered here in order to examine whether the highest $\rm T_{var}$ has indeed been estimated for each source.

The majority of flares with rise times $<14$ days have lower amplitudes than the median flare amplitude of the entire light curve suggesting that OVRO's three day sampling allowed us to resolve all the major events within the time span of the observations used in this work. Although low amplitude variability is still possible on shorter timescales it can be adequately described by a stochastic background process. Intra-day variability has been found in a handful of the brightest radio sources so it could be interesting to observe sources at an even faster cadence than 3 days, however, such fast variations are often attributed to interstellar scintillation and not to intrinsic processes.

Our results show a significant difference between the  $\rm T_{var}$ distributions of {\it Fermi} detected sources and non-detected sources. A similar result was obtained by \cite{Kovalev2009} when comparing the median VLBI brightness temperature of {\it Fermi} detected sources and non-detected sources from the MOJAVE survey.  A more in depth comparison of their flaring properties showed significant differences between {\it Fermi} detected and non-detected sources with the former showing on average faster flares with higher amplitudes. A comparison of the radio flux-density distributions of {\it Fermi} detected and non-detected sources in \cite{Liodakis2017-IV} also showed that $\gamma$-ray loud sources are systematically more variable and have higher flaring ratios (ratio of the flaring to quiescent mean flux densities) than $\gamma$-ray quiet sources. Our findings extend that result showing that both the variability and flaring properties in radio are connected to the $\gamma$-ray activity. This would suggest that the underlying mechanism in the jet that would cause the higher and more energetic flares in radio is, at least in part, also responsible for the $\gamma$-ray emission.

Using population models and the $\rm T_{var}$ estimates, we were able to effectively constrain the equipartition brightness temperature to  $\rm \langle T_{eq}\rangle=2.78\times10^{11}$K ($\pm26\%$). Previous attempts on constraining the limiting intrinsic brightness temperature had either estimated $\rm T_{eq}$ to be between $10^{10}-10^{11}$K \citep{Readhead1994,Lahteenmaki1999-II,Cohen2003} or most recently constrained it to $\rm T_{eq}>2\times10^{11}$K \citep{Homan2006}. The very high cadence of the OVRO program allowed us to resolve even the fastest events pushing the limit of the highest estimated  $ T_{\rm var}$ and hence provide the strongest constraints on $\rm T_{eq}$. Interestingly our results are consistent with the theoretical expectations for the limiting brightness temperature for incoherent synchrotron sources due to magnetization effects ($\sim3\times10^{11}$K, \citealp{Singal1986}). Although our results are model dependent, the fact that they are in agreement with both observational \citep{Homan2006} and theoretical \citep{Singal1986} expectations for blazar jets gives us confidence in our analysis.

Based on the results of the $\rm T_{eq}$ optimization, we estimated $\rm \delta_{var}$ and its uncertainty for the sources in our sample, significantly increasing the number of available $\rm \delta_{var}$ estimates in the literature. As expected, blazars are highly beamed sources with larger on average $\rm \delta_{var}$ than either radio galaxies or unclassified sources (median $\approx11$ for blazars compared to median of $\approx5$ for radio galaxies and unclassified sources). Surprisingly, we do not detect any significant difference between the BL Lacs and FSRQs populations contrary to the current consensus suggesting that FSRQs are more beamed than BL Lacs (H09, L17). This of course could be due to the selection of $\rm T_{eq}$ to be the same for all populations. If the BL Lacs were allowed to have the very high $\rm T_{eq}$ found in the above analysis (although not compatible with our current understanding of jet processes), it would lower their $\rm \delta_{var}$ estimates by a factor of $\sim2.5$. A more plausible explanation for this discrepancy could be that previous monitoring programs (with slower cadence than OVRO's) were not able to resolve the BL Lac flares evolving on faster timescales (see discussion in section \ref{TB_var}), but were still able to detect the slower evolving FSRQ flares. In such a case, it is only natural that FSRQs would show higher $\rm T_{\rm var}$ and hence larger $\delta_{\rm var}$ than BL Lacs. Contrary to previous monitoring programs, OVRO's fast cadence allowed us to detect all prominent flares in both BL Lacs and FSRQs.

Although there is no significant difference between the viewing angle distributions of the blazar classes (not surprising if the sources are uniformly distributed and randomly oriented) FSRQs host faster jets than BL Lacs. This could help explain differences between the two populations, or at least, differences between FSRQs and radio bright BL Lacs. As expected {\it Fermi} detected sources have on average faster jets pointed at smaller angles towards our line of sight than {\it Fermi} non-detected sources. Then the relativistic effects could also be partly responsible for the detected (or not) $\gamma$-ray emission in addition to radio variability and flaring properties (see also \citealp{Lister2009-III}). It should be noted that while the flux-density variations used to estimate the brightness temperature and Doppler factors originate predominately in the radio core of the jet and are believed to be related to ejections of new radio components (e.g., \citealp{Savolainen2002}), apparent velocities are measured downstream from the core. Given that the observations used in this work and apparent velocity measurements from the MOJAVE program are taken at the same radio frequency (15~GHz, and thus probe the same region for a given source), we do not expect significant changes in the velocity of the jet over short distances. However, since both accelerating and decelerating jet components have been measured at 15~GHz (e.g., \citealp{Homan2015}), our results for the Lorentz factors and viewing angles in sources that show large velocity gradients should be treated as limits. Additionally, to derive the $\rm \Gamma_{var}$, $\rm \theta_{var}$ estimates we have used the maximum observed apparent velocity in each jet. Although it has been shown that the radio components of individual jets are ejected at similar velocities \citep{Lister2013}, using a different measure for $\beta_\mathrm{app}$ (e.g., mean, median) could result in differences in the $\rm \Gamma_{var}$ and $\rm \theta_{var}$ estimates.

A comparison with previous attempts (H09, L17) in estimating $\rm \delta_{var}$ showed that (after accounting for the different assumed $\rm T_{eq}$) the estimates from this work are systematically higher with only 20\% of the common estimates to be consistent within 1$\sigma$. This is not surprising given the faster cadence of the OVRO survey. On the other hand, comparing our estimates with those from J17 derived using a different approach independent of equipartition showed that 30.5\% of the sources in the J17 sample are consistent with the estimates from this work. This agreement allows us to place strong constraints on the $\rm \delta$ estimates for these sources.

\acknowledgments
We thank the anonymous referee for comments and suggestions that helped improve this work. This research has made use of data from the OVRO 40-m monitoring program \citep{Richards2011} which is supported in part by NASA grants NNX08AW31G, NNX11A043G, and NNX14AQ89G and NSF grants AST-0808050 and AST-1109911. This research has made use of data from the MOJAVE database that is maintained by the MOJAVE team \citep{Lister2018}. This research has made use of the SIMBAD database,
operated at CDS, Strasbourg, France \citep{Wenger2000}. This research has made use of the NASA/IPAC Extragalactic Database (NED), which is operated by the Jet Propulsion Laboratory, California Institute of Technology, under contract with the National Aeronautics and Space Administration.
\facilities{OVRO}
\software{Magnetron \citep{Huppenkothen2015}, DNest4 \citep{Brewer2016}, Numpy \citep{Walt2011}, Scipy \citep{Jones2001}.}

\bibliographystyle{apj}
% Use the LaTeX power, use bibtex properly.
\bibliography{bibliography} %graphy.bib}%,bibliography_export.bib}


\end{document}